\RequirePackage[running]{lineno}

\documentclass[aps,prl,twocolumn,nofootinbib,superscriptaddress,letterpaper,amsmath,amssymb]{revtex4}
\pdfoutput=1

\usepackage{graphicx}  
\usepackage{dcolumn}   
\usepackage{bbm}        
\usepackage{amsmath}
\usepackage{amsfonts}
\usepackage{amssymb}   
\usepackage{epstopdf}
\usepackage{slashed}
\usepackage{hyperref}
\usepackage{multirow}
\usepackage{color}
\usepackage{float}
\usepackage{ragged2e}
\usepackage{subcaption}
\DeclareCaptionJustification{justified}{\justifying}
\usepackage{verbatim}
\usepackage{xspace}

\usepackage[compat=1.1.0]{tikz-feynman}

\def\gsim{\lower0.5ex\hbox{$\:\buildrel >\over\sim\:$}}
\def\lsim{\lower0.5ex\hbox{$\:\buildrel <\over\sim\:$}}

\newcommand{\rev}[1]{{ #1}}

\newcommand{\met}{$p_{\rm T}^{\rm miss}$\xspace}
\newcommand{\be}{\begin{equation}}
\newcommand{\ee}{\end{equation}}
\newcommand{\bea}{\begin{eqnarray}}
\newcommand{\eea}{\end{eqnarray}}

\newcommand{\nbox}{{\,\lower0.9pt\vbox{\hrule \hbox{\vrule height 0.2 cm
\hskip 0.2 cm \vrule height 0.2 cm}\hrule}\,}}

\def\to{\rightarrow}

\newskip\zatskip \zatskip=0pt plus0pt minus0pt
\def\matth{\mathsurround=0pt}
\def\lsim{\mathrel{\mathpalette\atversim<}}
\def\gsim{\mathrel{\mathpalette\atversim>}}
\def\sigv{\ifmmode \langle\sigma v\rangle\else $\langle\sigma v\rangle$\fi}
\newskip\zatskip \zatskip=0pt plus0pt minus0pt
\def\matth{\mathsurround=0pt}
\def\lsim{\mathrel{\mathpalette\atversim<}}
\def\gsim{\mathrel{\mathpalette\atversim>}}
\def\atversim#1#2{\lower0.7ex\vbox{\baselineskip\zatskip\lineskip\zatskip
  \lineskiplimit
  0pt\ialign{$\matth#1\hfil##\hfil$\crcr#2\crcr\sim\crcr}}}

\begin{document}

\thispagestyle{empty}

\vspace{0.5in}

\title{Hadronic Mono-$W'$ Probes of Dark Matter at Colliders}
\author{Ryan Holder}
\affiliation{Department of Physics \& Astronomy, University of California, Irvine, CA}
\author{John Reddick}
\affiliation{Department of Physics, Irvine Valley Community College, Irvine, CA}
\author{Matteo Cremonesi}
\affiliation{Department of Physics, Carnegie Mellon University, Pittsburgh, PA}
\author{Doug Berry}
\affiliation{Particle Physics Division, Fermi National Accelerator Laboratory, Batavia, IL}
\author{Kun Cheng}
\affiliation{School of Physics, Peking University, Beijing, China}
\affiliation{PITT PACC, Department of Physics and Astronomy, University of Pittsburgh, Pittsburgh, PA}
\author{Matthew Low}
\affiliation{PITT PACC, Department of Physics and Astronomy, University of Pittsburgh, Pittsburgh, PA}
\author{Tim M.P. Tait}
\author{Daniel Whiteson}
\affiliation{Department of Physics \& Astronomy, University of California, Irvine, CA}

\begin{abstract}
Particle collisions at the energy frontier can probe the nature of invisible dark matter via production in association with recoiling visible objects. We propose a new potential production mode, in which dark matter is produced by the decay of a heavy dark Higgs boson radiated from a heavy $W'$ boson. In such a model, motivated by left-right symmetric theories, dark matter would not be pair produced in association with other recoiling objects due to its lack of direct coupling to quarks or gluons. We study the hadronic decay mode via $W'\rightarrow tb$ and estimate the LHC exclusion sensitivity at 95\% confidence level to be $10^2-10^5$ fb for $W'$ boson masses between 250 and 1750 GeV.
\end{abstract}
\maketitle

\section{Introduction}

Though dark matter (DM) represents the majority of the matter density of the universe, its particle nature remains a mystery. A variety of DM candidates have been considered, many with connections to the electroweak symmetry-breaking scale~\cite{Jungman:1995df, Bertone:2004pz}, which would also explain  the observed relic density~\cite{Scherrer:1985zt}. A robust program of dedicated experiments search for DM interactions~\cite{LUX:2016ggv,XENON:2023cxc,PandaX-II:2017hlx,Bertone:2018krk}, which have not yet detected a signal. 

At particle colliders, searches for DM production focus on the visible recoil $X$ from the invisible DM, which leaves missing transverse momentum, \met. Cases where $X$ is a SM particle such as a quark or a gluon~\cite{Beltran:2010ww,Fox:2011pm,Goodman:2010ku,Rajaraman:2011wf,Aad:2015zva,Khachatryan:2014rra}, a $W$ boson~\cite{Bai:2012xg,Aad:2013oja,Khachatryan:2014tva}, a $Z$ boson~\cite{Bell:2012rg,Carpenter:2012rg,Aad:2014vka}, a Higgs boson~\cite{Carpenter:2013xra,Berlin:2014cfa}, a photon~\cite{Fox:2011pm,Khachatryan:2014rwa,Aad:2014tda}, or a non-SM particle such as a $Z'$ boson~\cite{Autran:2015mfa}, a leptonically-decaying $W'$ boson~\cite{ATLAS:2014wra}, or a heavy quark~\cite{Lin:2013sca,Haisch:2015ioa,CMS:2014pvf,Aad:2014vea} have been considered. For a review of simplified models for DM at the LHC, see Refs.~\cite{Abdallah:2014hon,Malik:2014ggr}.

In this paper we describe a new search mode, in which DM recoils against a heavy $W'$ boson that decays to a hadronically decaying top quark and a $b$ quark (referred to as the $tb$ final state). This mode provides a statistically independent and theoretically distinct probe of DM production from other \met$+X$ searches.

In addition to probing DM, this channel also probes the $W'$ boson itself. While the LHC already sets very stringent limits on high-mass $W'$ bosons~\cite{ATLAS:2023ibb,CMS:2021mux}, searches at lower masses \rev{($m_{W'}\lessapprox 1$ TeV)} are more challenging due to the stringent trigger requirements on the decay products of the $W'$ boson. The recoiling DM allows  \met-based triggers to be used, which opens up the possibility to push $W'$ boson searches to lower masses.

The paper is organized as follows. A model of DM production in association with a $W'$ boson is presented. Selection and reconstruction strategies are proposed and the expected sensitivity of the LHC dataset is described. The final section puts the expected sensitivity in experimental and theoretical context.

\section{Model}
\label{sec:model}

\begin{figure} [tb]
    \centering
    \begin{tikzpicture}
    \begin{feynman}
        \vertex (a) ;
        \vertex [above left=of a](i1){\(q\)};
        \vertex [below left=of a](i2){\(\bar q\)};
        \vertex [right=of a] (b);
        \vertex [above right=of b] (f1) {\(h_{D}\)};
        \vertex [below right=of b] (c);
        \vertex [above right=of c] (f2) {\(\overline t\)};
        \vertex [below right=of c] (f3) {\(b\)};
        \diagram* {
         (i1) -- [fermion] (a) -- [fermion] (i2),
        (a) -- [photon, edge label'=\(W'\)] (b) -- [ghost] (f1),
        (b) -- [boson, edge label'=\(W'\)] (c),
        (c) -- [anti fermion] (f2),
        (c) -- [fermion] (f3),
    };
    \end{feynman}
    \end{tikzpicture}
    \caption{Feynman diagram describing the  production of a heavy  $W'$ boson recoiling against a dark Higgs boson ($h_D$) and decaying to a top and a bottom quark. If the $s$-channel $W'$ boson is virtual (real), the decay is 2-body (3-body).}
    \label{fig:decay-diagram}
\end{figure}
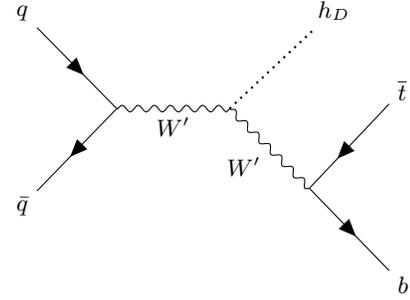

We present a model of a heavy $W'$ boson that can be produced in association with invisible DM particles via the radiation of a dark Higgs boson, which decays to DM particles. Such a model would not produce a signature in other \met$+X$ search modes.\footnote{As will be discussed, the $Z'$ boson is assumed to be substantially more massive than the $W'$ boson.  Additionally, we require that the DM be pair produced.}

The $W'$ boson is a new gauge boson that commonly arises in models of new physics, such as extended gauge theories~\cite{Mohapatra:1974gc,Mohapatra:1974hk,Georgi:1989ic,Chivukula:1994mn,Malkawi:1996fs} or composite Higgs~\cite{Panico:2015jxa}. In this work, we consider extending the electroweak gauge group $SU(2)_L \times U(1)_Y$ to $SU(2)_L \times SU(2)_R \times U(1)_{B-L}$ typically  known as the left-right symmetric model~\cite{Mohapatra:1974gc,Mohapatra:1974hk}.

In this model, the $SU(2)_R \times U(1)_{B-L}$ symmetry is broken to $U(1)_Y$ which results in one new massive charged gauge boson, the $W'$ boson, and one new massive neutral gauge boson, the $Z'$ boson. This symmetry breaking is accomplished through the vacuum expectation value of an additional scalar multiplet.

The fermion content of the theory is the same as the SM, with the addition of a right-handed neutrino, $N_R$. The gauge representations of the fermions under $SU(2)_L \times SU(2)_R \times U(1)_{B-L}$ are:
\begin{equation}
\begin{array}{ll}
Q_{L, i}=\left(\begin{array}{c}
u_L \\
d_L
\end{array}\right)_i:\left( \mathbf{2}, \mathbf{1}, \frac{1}{3}\right), & Q_{R, i}=\left(\begin{array}{c}
u_R \\
d_R
\end{array}\right)_i:\left( \mathbf{1}, \mathbf{2}, \frac{1}{3}\right), \\
\psi_{L, i}=\left(\begin{array}{c}
\nu_L \\
e_L
\end{array}\right)_i:( \mathbf{2}, \mathbf{1},-1), & \psi_{R, i}=\left(\begin{array}{c}
N_R \\
e_R
\end{array}\right)_i:( \mathbf{1}, \mathbf{2},-1).
\end{array}
\end{equation}
The scalar content consists of a bi-doublet $\phi$, which contains the SM Higgs doublet, and an $SU(2)_R$ triplet $\Delta_R$:
\begin{align}
  \phi&=\begin{pmatrix}
    \phi_1^0 & \phi_1^+ \\
    \phi_2^- & \phi_2^0
    \end{pmatrix}:  (\mathbf{2,2},0),\\
    \Delta_R&=\begin{pmatrix}
        \delta_{R}^+/\sqrt{2} & \delta_{R}^{++} \\
        \delta_{R}^0 & -\delta_{R}^+/\sqrt{2}
    \end{pmatrix}: (\mathbf{1,3},2).
\end{align}
We assume the potentials are engineered such that these scalars have the vacuum expectation values:
\begin{equation}
  \langle\phi\rangle=\frac{1}{\sqrt{2}}\left(\begin{array}{cc}
    \kappa_1 & 0 \\
    0 & \kappa_2
  \end{array}\right), \quad
  \left\langle\Delta_R\right\rangle=\frac{1}{\sqrt{2}}\left(\begin{array}{cc}
    0 & 0 \\
    v_R & 0
  \end{array}\right),
\end{equation}
where $v^2=\kappa_1^2+\kappa_2^2=(246~{\rm GeV})^2$ and $v_R$ is a free parameter.

After accounting for the states that give mass to the $W'$ and $Z'$ bosons, the triplet $\Delta_R$ contains one neutral state, one charged state, and one doubly-charged state. We call the neutral state a dark Higgs boson $(h_D)$ due to its lack of direct interactions with the SM quarks, preventing it from being produced in an $s$-channel process at the LHC. It is given by $\delta_R^0=(v_R+h_D)/\sqrt{2}$.

For simplicity we assume no mixing between scalars. In principle the SM Higgs boson and the $h_D$ boson can mix, however, experimentally a non-zero mixing would still need to be small, $\lesssim {\cal O}(10\%)$~\cite{ATLAS:2022vkf,CMS:2018uag}.

In this model the mass of the $W'$ is
\begin{equation}
\label{eq:wpmass}
M_{W'} = \frac{g_R}{2} \sqrt{v^2+2v_R^2},
\end{equation}
where $g_R$ is the gauge coupling of $SU(2)_R$. Using Eq.~\eqref{eq:wpmass}, the mass of the $W'$ boson $(m_{W'})$ can be specified, rather than the value of $v_R$, such that the relevant parameter space of this model is $m_{W'}$ and $g_R$. The gauge coupling of $U(1)_{B-L}$ is determined by the choice of $g_R$ since $SU(2)_R \times U(1)_{B-L} \to U(1)_Y$.

The hadronic decay of $W' \to tb$ is mediated by the interaction
\begin{equation}
\mathcal{L}
= \frac{g_R}{\sqrt{2}}(\bar u_R \gamma^\mu d_R) W'^+_\mu + h.c.,
\end{equation}
while the production cross section of $pp\to W' \to h_D t b$ is proportional to $(g_R^3 v_R)^2$. When $m_{W'}$ is fixed and $v_R\gg v$, the cross-section scaling becomes proportional to $g_R^4$.

The mass of the $Z'$ boson in this model is 
\begin{equation}
m_{Z'} = \sqrt{(g_R^2+g_{BL}^2)v_R^2 + \frac{g_R^4}{4(g_R^2+g_{BL}^2)} v^2},
\end{equation}
where $g_{BL}$ is the gauge coupling of $U(1)_{B-L}$. The $Z'$ boson couples to leptons, which would be visible unless the $Z'$ boson is heavy enough to avoid experimental bounds. When $m_{W'} \approx 800~{\rm GeV}$ and $g_{BL} \gtrsim 2.5$, the mass of the $Z'$ boson $(m_{Z'}) \gtrsim 7~{\rm TeV}$.

In this minimal version of a left-right model, the $h_D$ boson can dominantly decay to right-handed neutrinos $N_R$. If these are sufficiently light, less than of order keV, then they could comprise the majority of the DM in the universe~\cite{Drewes:2013gca}.  More generally, the DM could be any new stable particle.  Our search, like other collider searches, is agnostic to the identity of the DM.

\section{Experimental Sensitivity}

The model described above includes interactions which can generate a final state with a top quark, a bottom quark, and missing transverse momentum, see Fig.~\ref{fig:decay-diagram}. We estimate the sensitivity of the LHC dataset to these hypothetical signals using samples of simulated $pp$ collisions at \mbox{$\sqrt{s}=13$ TeV} with an integrated luminosity of 300 fb$^{-1}$.

Simulated signal and background samples are used to model the reconstruction of the $W'$ boson candidates, estimate selection efficiencies, and expected signal and background yields. Collisions and decays are simulated with {\sc Madgraph5} v3.4.1 ~\cite{madgraph}, and {\sc Pythia} v8.306~\cite{pythia} is used for fragmentation and hadronization. The model for the $W'$ boson was adapted in FeynRules~\cite{Alloul:2013bka} from Ref.~\cite{Roitgrund:2014zka}. The detector response is simulated with {\sc Delphes} v3.5.0~\cite{delphes} using the standard CMS card, extended to include an additional reconstruction of wide-cone jets, and {\sc root} version 6.26\/06 \cite{ROOT}. 

Selected narrow-cone (wide-cone) jets are clustered using the anti-$k_{\textrm{T}}$ algorithm~\cite{Cacciari:2008gp} with radius parameter $R = 0.4$ ($R=1.2$) using \textsc{FastJet 3.1.2}~\cite{Cacciari:2011ma} and are required to have $p_\textrm{T}\geq20$ GeV and $0\leq|\eta|\leq2.5$. Wide-cone jets with mass within $[50,110]$ ($[125,225]$) GeV are tagged as $W$-boson (top-quark) jets. Events are required to have no reconstructed isolated photons, muons, or electrons with $p_\textrm{T}\geq10$ GeV and $|\eta|\leq2.5$; \rev{ isolation requires that less than 12\% (25\%) of the $p_\textrm{T}$ of the electron or photon (muon) be deposited in a cone with $\Delta R < 0.5$ centered on the particle}. To satisfy a trigger requirement and suppress backgrounds, events must have at least 200 GeV of \met.

Candidate $W'$ bosons are reconstructed in one of three approaches: 

\begin{itemize}
    \item $t+b$: one top-tagged, $b$-tagged wide-cone jet and a $b$-tagged narrow-cone jet
    \item $W+b+b$: one $W$-tagged, un-$b$-tagged wide-cone jet and two $b$-tagged narrow-cone jets
    \item $jj+b+b$: two un-$b$-tagged narrow-cone jets and two $b$-tagged narrow-cone jets
\end{itemize}

\begin{figure}
    \centering
    \includegraphics[width=0.45\textwidth]{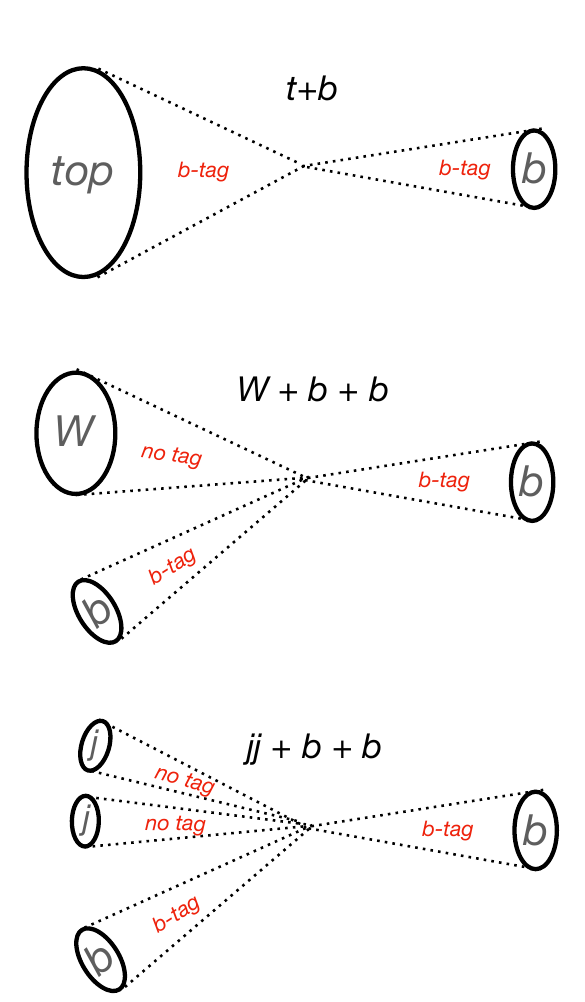}
    \caption{Three possible $W'$ boson reconstruction strategies, using wide-cone and narrow-cone jets, which can be $b$-tagged, $W$-tagged or top-tagged. See text for details.}
    \label{fig:topo}
\end{figure}

\begin{figure}
    \centering
    \includegraphics[width=0.45\textwidth]{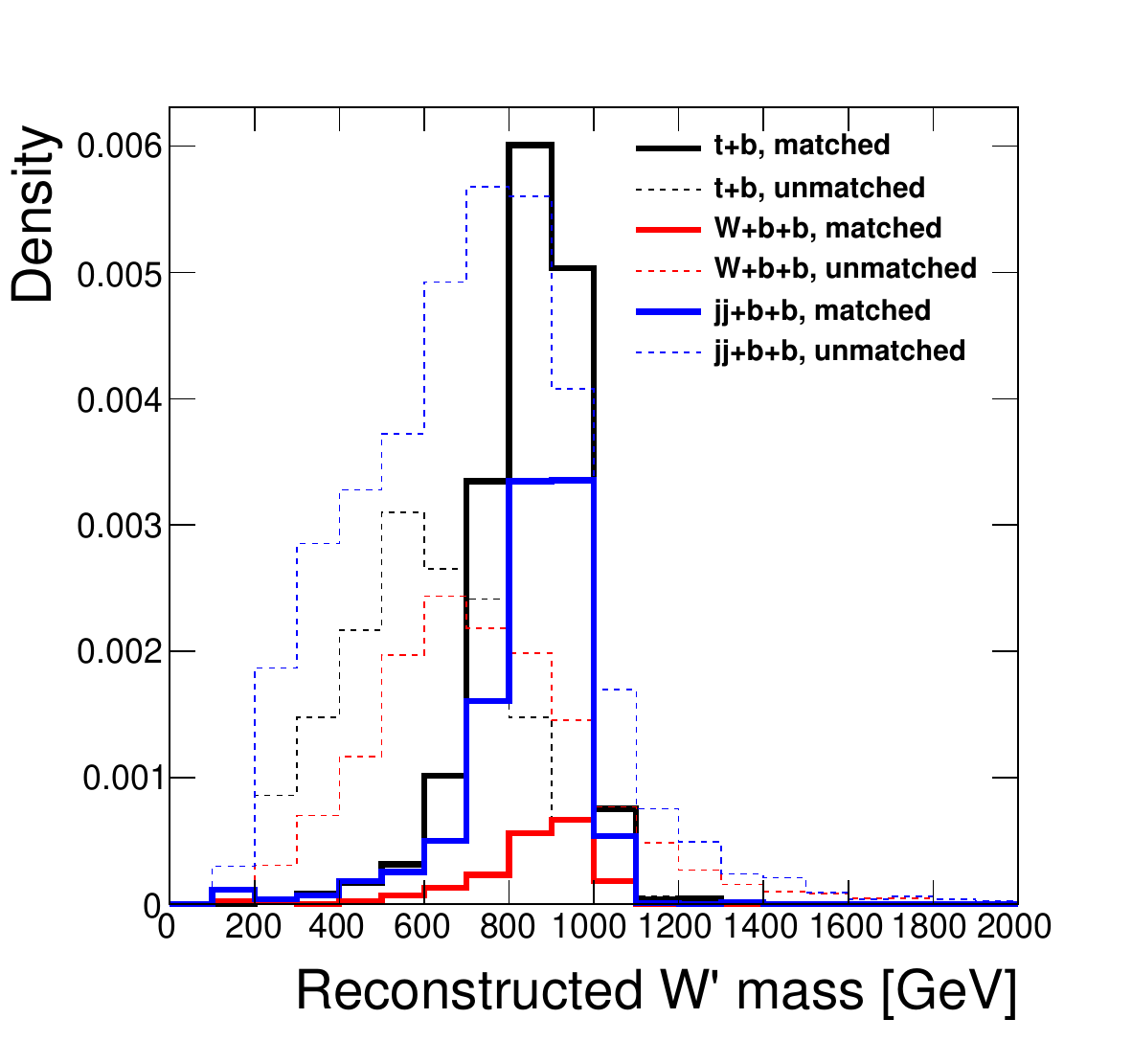}
    \includegraphics[width=0.45\textwidth]{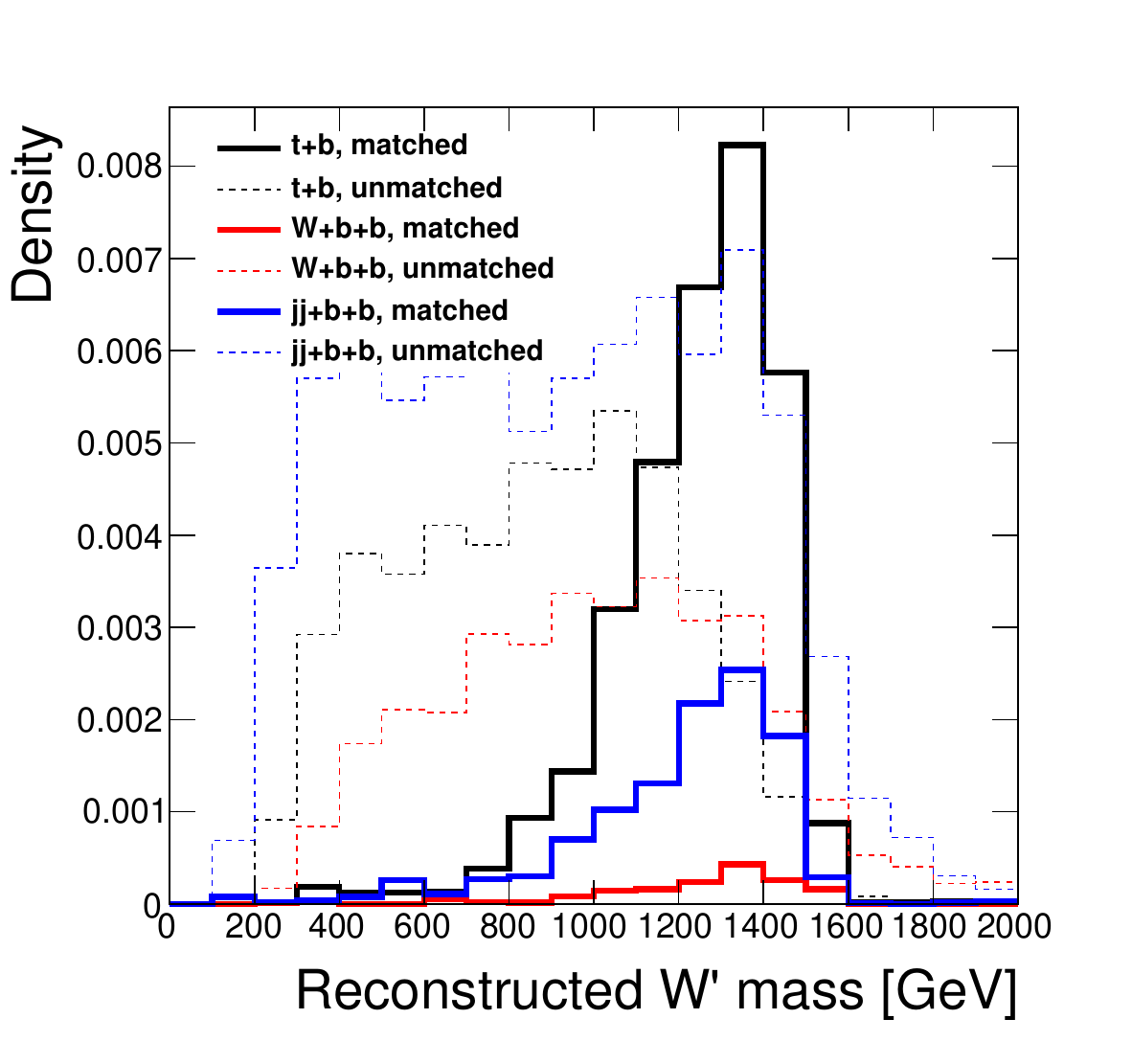}
    \caption{Top (bottom): Distribution of the reconstructed $W'$ boson candidate mass in simulated events with $m_W'=1000$ (1500) GeV, for each of the three reconstruction strategies (see Fig~\ref{fig:topo}), where the selected objects are angled-matched ($\Delta R<0.4$) and -unmatched ($\Delta R>0.4$) to the correct parton-level objects} 
    \label{fig:reco}
\end{figure}

\begin{figure}
    \centering
    \includegraphics[width=0.45\textwidth]{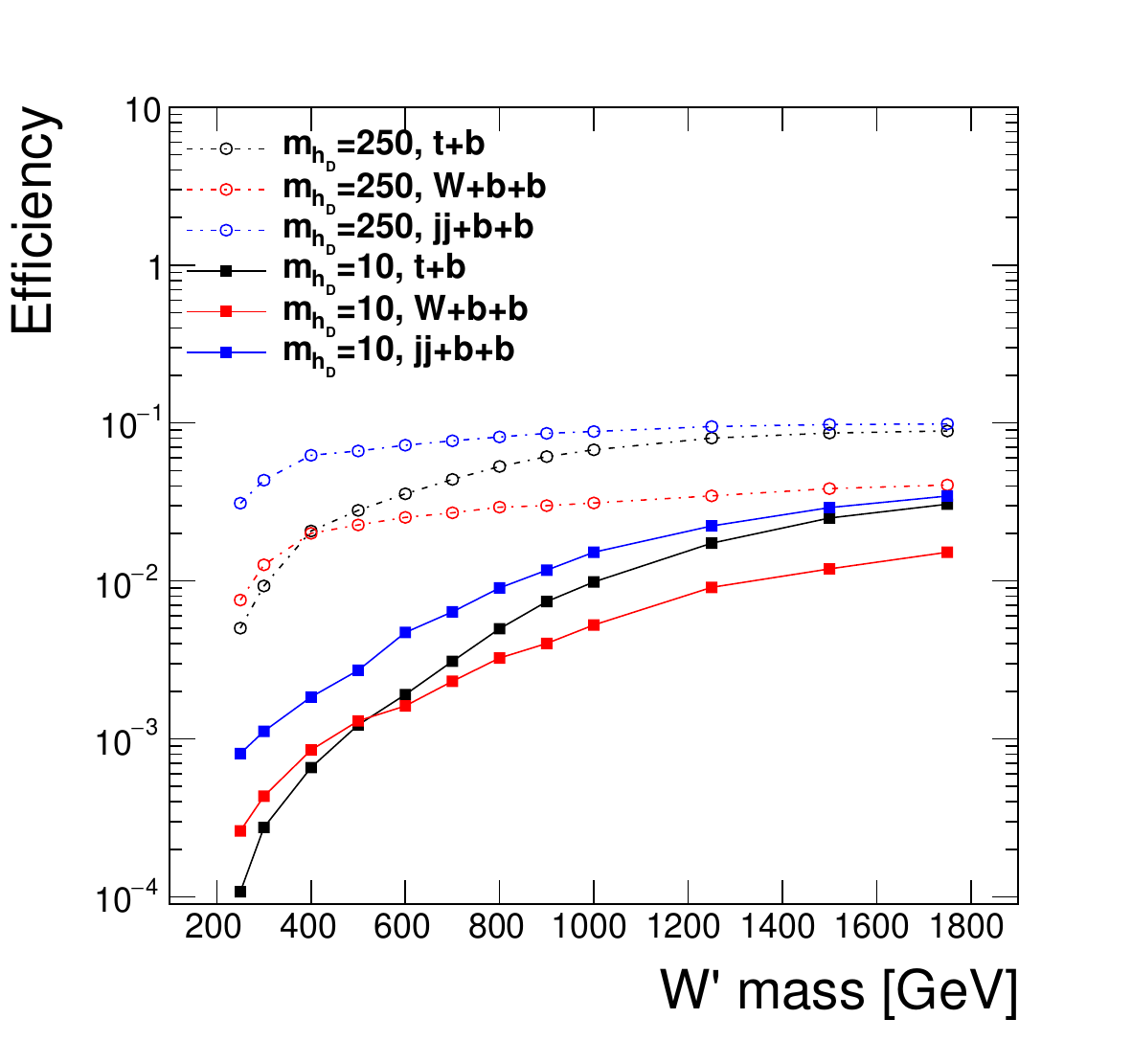}
    \caption{Efficiency of the selection in each approach ($t+b$, $W+b+b$, $jj+b+b$, see Fig~\ref{fig:topo}) as a function of the $W'$ boson mass, for two choices of the \mbox{dark Higgs boson mass $(m_{h_{D}})$}.}
    \label{fig:eff}
\end{figure}

The three approaches are illustrated in Fig.~\ref{fig:topo}. If several reconstruction approaches are available for a single event, preference is given to $t+b$ and then $W+b+b$. If several jets are available within one approach, preference is given to the jets that minimize the difference between the reconstructed and known top-quark and $W$-boson masses. Distributions of reconstructed $W'$ boson candidate masses are shown in Fig.~\ref{fig:reco} for two choices of $W'$ boson mass, and the selection efficiency is shown in Fig~\ref{fig:eff}. Collider searches for signals with this event topology typically focus on the hadronic decay of the $W$ boson~\cite{CMS:2018rkg,ATLAS:2020lks} as it has a larger branching fraction and its decay can be efficiently reconstructed using large-radius jets and state-of-the-art tagging algorithms. The background from multijet events can be accurately modelled using data driven methods.

\begin{figure}
    \centering
        \includegraphics[width=0.45\textwidth]{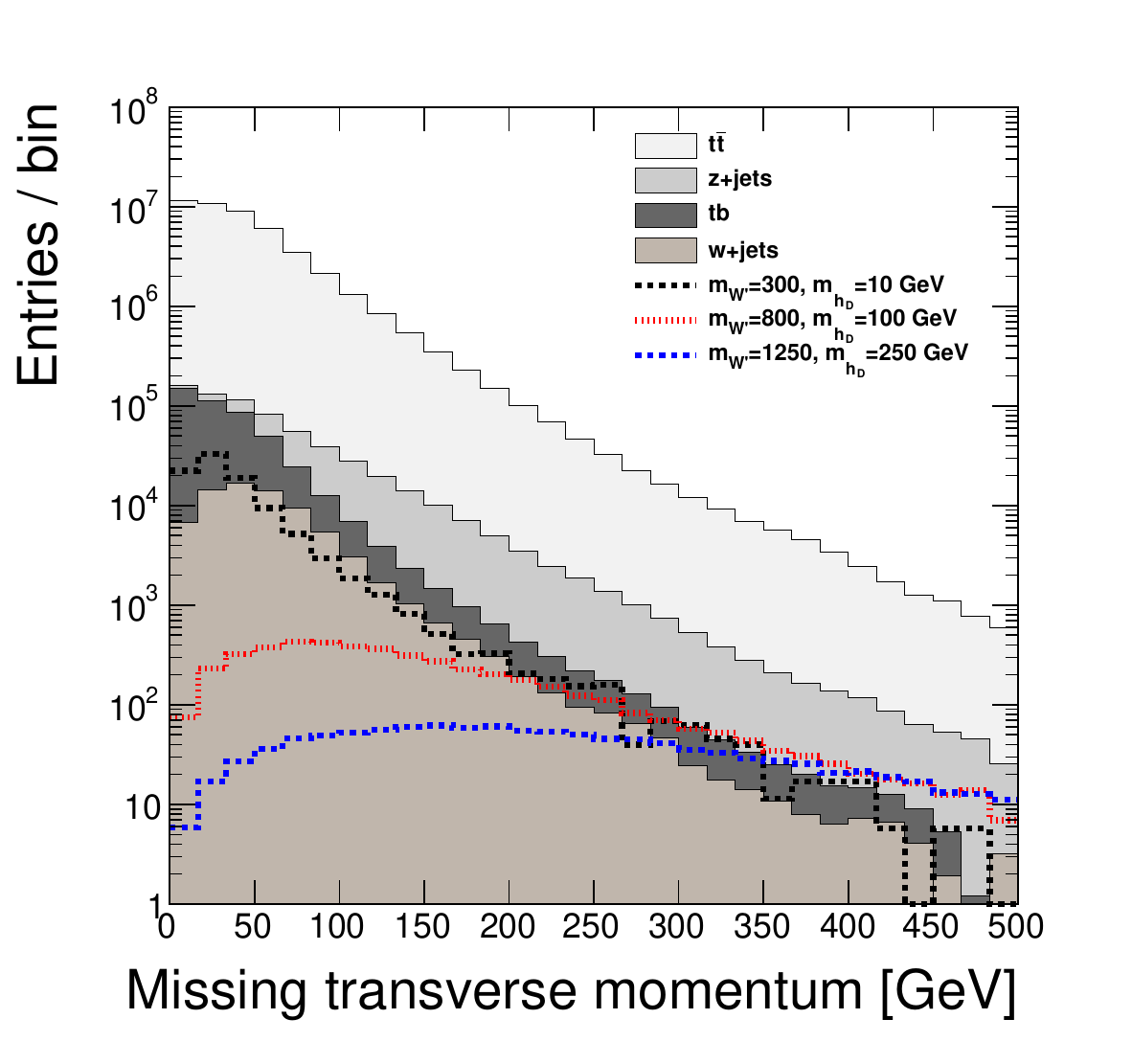}
    \includegraphics[width=0.45\textwidth]{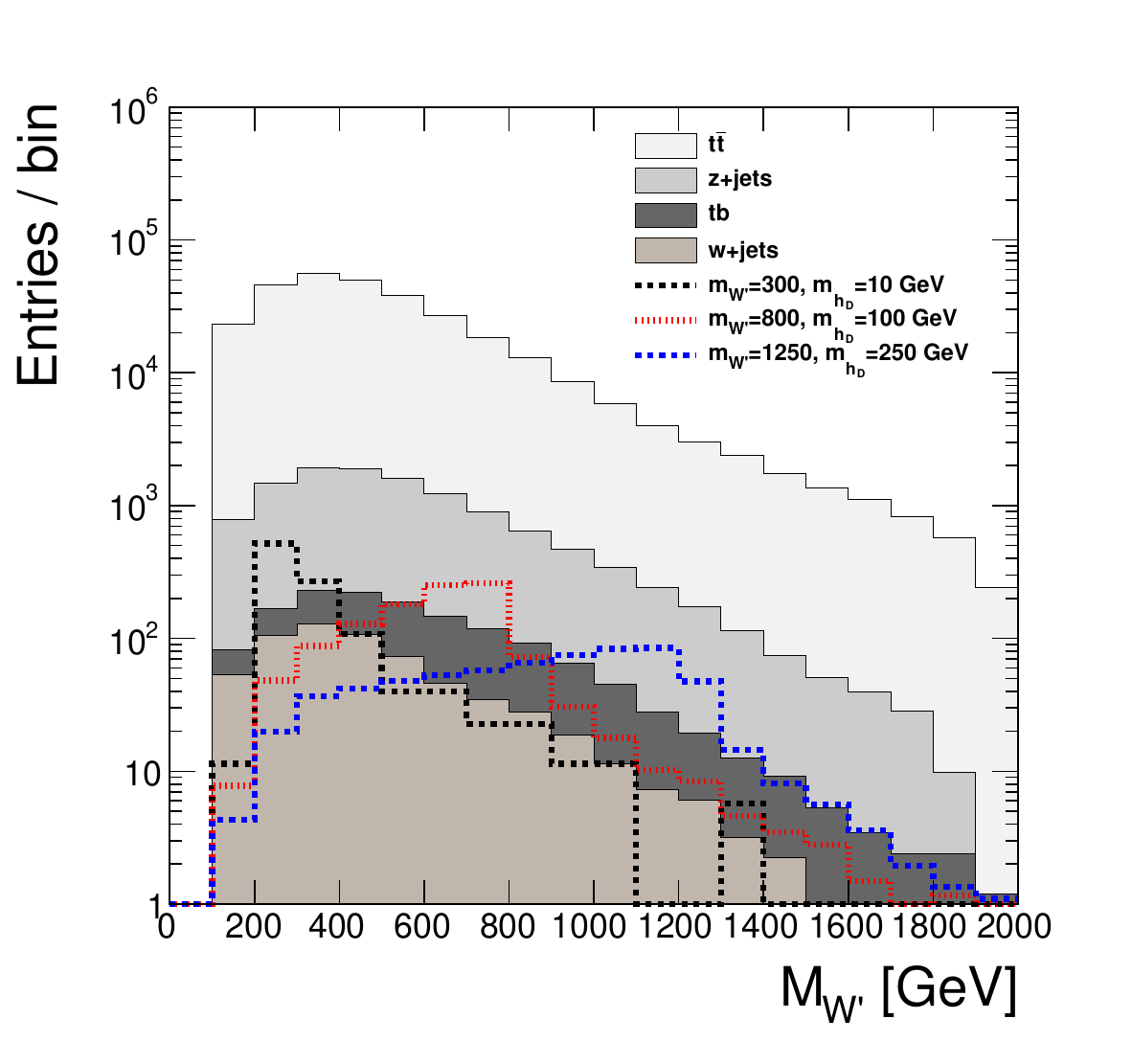}
    \caption{Top: Distribution of the missing transverse momentum (\met) for the expected background and selected signals normalized to an integrated luminosity of 300 fb$^{-1}$ after all requirements other than \met$>200$ GeV are met. Bottom: Distribution of the reconstructed $W'$ boson 2-body candidate mass for the expected background and selected signals normalized to an integrated luminosity of 300 fb$^{-1}$ after the full selection.}
    \label{fig:sb}
\end{figure}

The dominant backgrounds are the production of top-quark pairs ($t\bar{t}$) or the production of a single top quark in association with a $b$ quark ($t\bar{b}$ or $\bar{t}b$). Additional backgrounds are due to production of a heavy vector boson ($W$ or $Z$ bosons), which decays invisibly or whose decay products are not reconstructed, in association with two $b$ quarks and two additional hard quarks or gluons. Radiation of additional gluons is modeled by {\sc Pythia}. Contributions from QCD multi-jet production is suppressed by the \met requirement. Distributions of the expected reconstructed $W'$ boson masses for the background and signal processes are shown in Fig~\ref{fig:sb} and the expected yields in 300 fb$^{-1}$ are shown in Table~\ref{tab:yields}.

\begin{table}[]
    \centering
     \caption{Expected yields in 300 fb$^{-1}$ of LHC data for background and signal $(W'\rightarrow tb)$ processes. Cross sections for backgrounds are at NLO in QCD~\cite{Alwall:2014hca}; cross sections for signal are set to the expected 95\% CL upper limit. The calculations are described in the text. Shown are the cross section ($\sigma$), the trigger and selection efficiency ($\varepsilon$), and the expected yield ($N$).}
    \label{tab:yields}
    \begin{tabular}{l|ccc}
    \hline\hline\\
    Process & $\sigma$ [fb] & $\varepsilon$ & $N$\\
    \hline
    $t\bar{t}$     & $6.74 \times 10^5$	& $1.42 \times 10^{-3}$&	$2.89\times 10^{5}$ \\
    $Z+b\bar{b},Z\rightarrow \nu\nu$	&$2.47\times 10^{5}$&	$1.42\times 10^{-4}$	&10560\\
    $t\bar{b}+\bar{t}b$&	$1.00\times 10^{4}$	&$2.7\times 10^{-4}$&	820\\
    $W^\pm+b\bar{b},W^\pm\rightarrow \ell^\pm\nu$	&$1.74\times 10^{5}$&	$1.2\times 10^{-5}$ & 620\\
\hline
$M_W'=300,M_{h_D}=10$	& 2280 &	0.0016	&1060\\
$M_W'=800,M_{h_D}=100$	&66&	0.056	&1120\\
$M_W'=1250,M_{h_D}=250$	&16.9&	0.129	&650\\
         \hline\hline
    \end{tabular}
   
\end{table}

\begin{figure}
    \centering
    \includegraphics[width=0.45\textwidth]{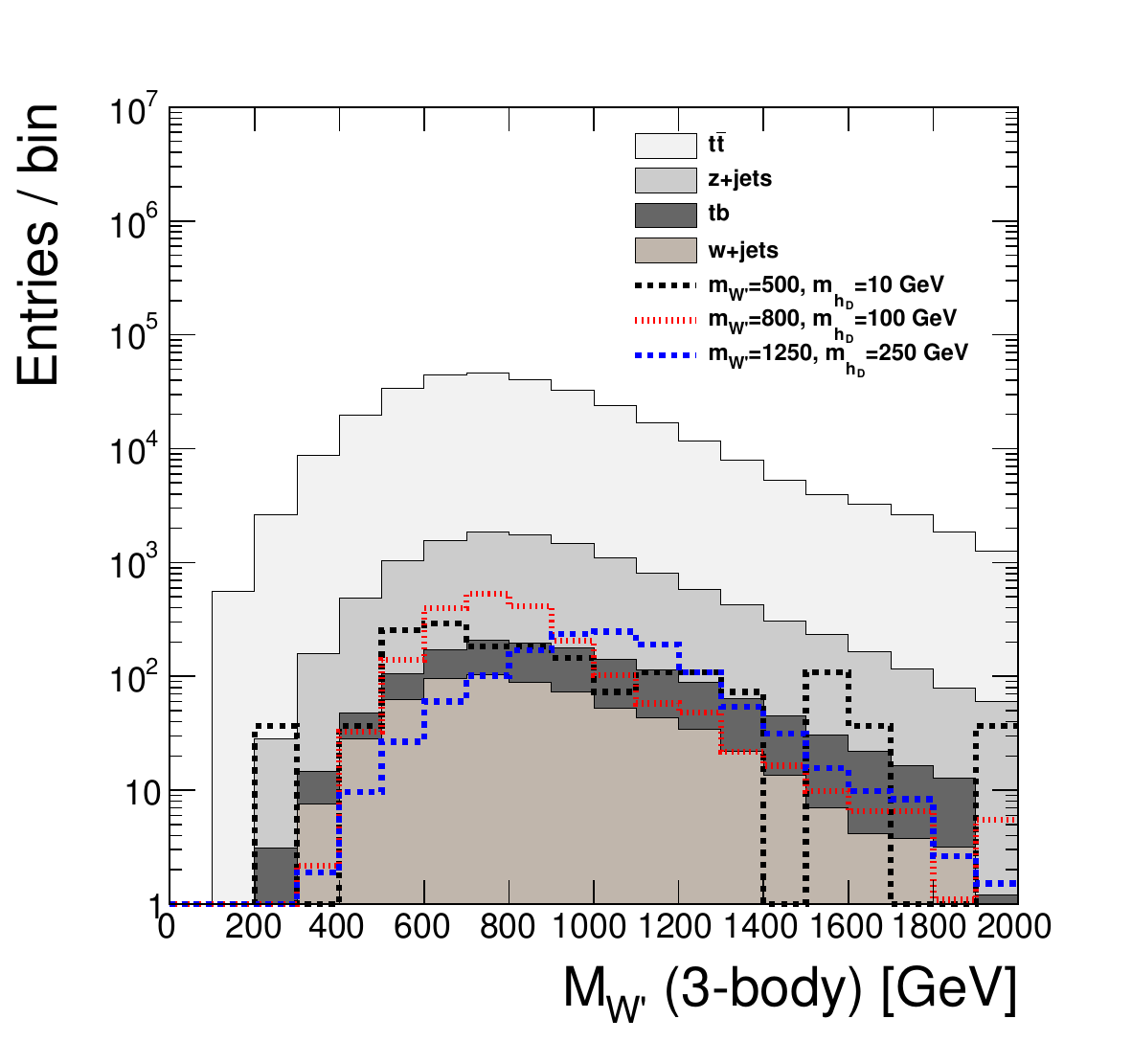}
    \caption{Distribution of the reconstructed $W'$ boson 3-body candidate mass for the expected background and selected signals normalized to an integrated luminosity of 300 fb$^{-1}$ after the full selection.}
    \label{fig:sbmet}
\end{figure}

We also consider the 3-body decay of the $W'$ boson, in which the $h_D$ boson is a $W'$ boson decay product rather than radiation from an on-shell $W'$ boson. Reconstruction of the $W'$ boson, in principle, requires knowledge of the invisible $h_D$ boson's four-momentum. We reconstruct the 3-body decay using the same techniques as for the 2-body decay, but with \met added to the $W'$ boson candidate as an estimate of the $h_D$ boson transverse momentum. No estimate is made of the longitudinal momentum of the $h_D$ boson. Distributions of expected background and signals are shown in Fig.~\ref{fig:sbmet}.

\begin{figure}
    \centering
   \includegraphics[width=0.45\textwidth]{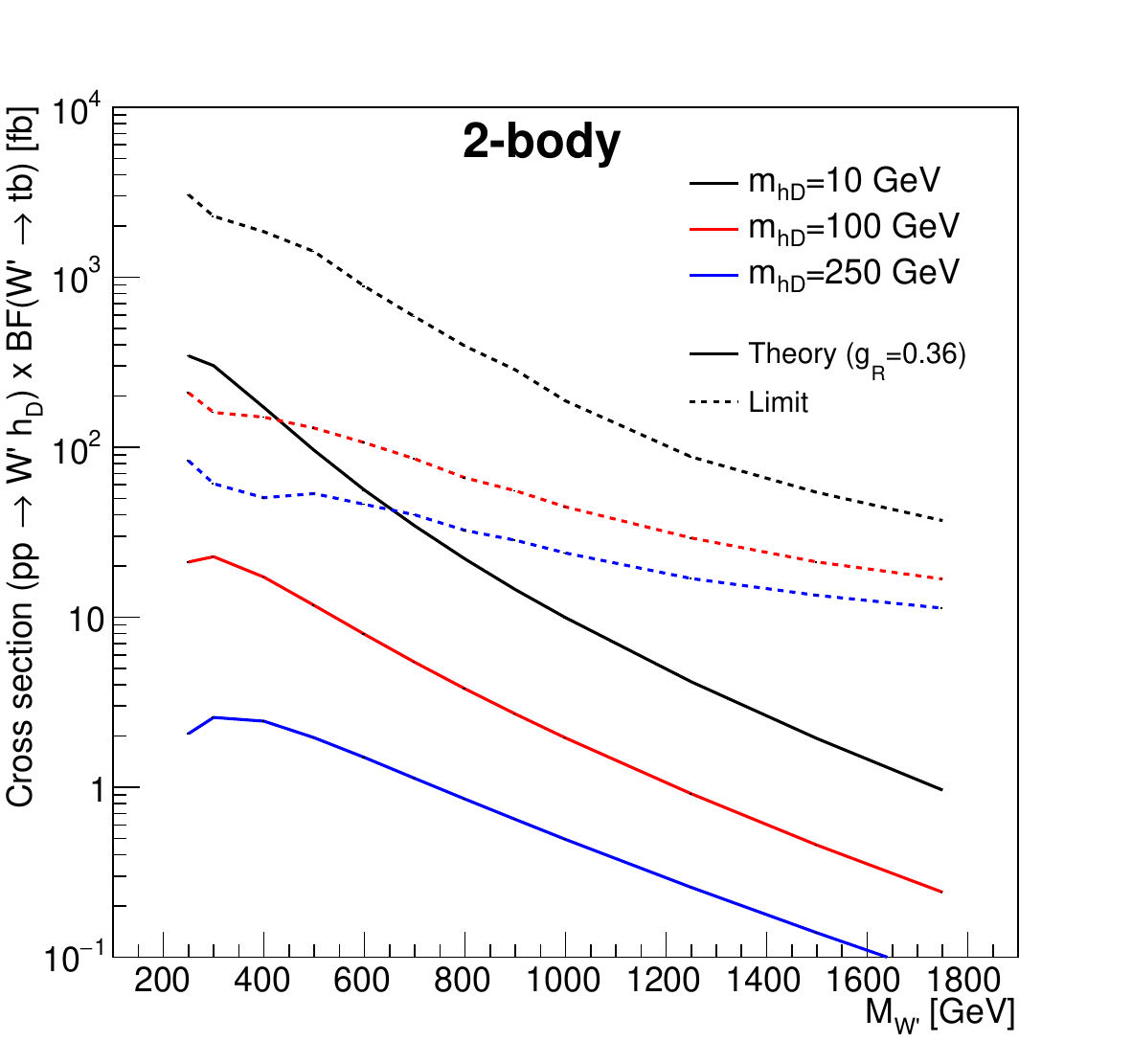}
      \includegraphics[width=0.45\textwidth]{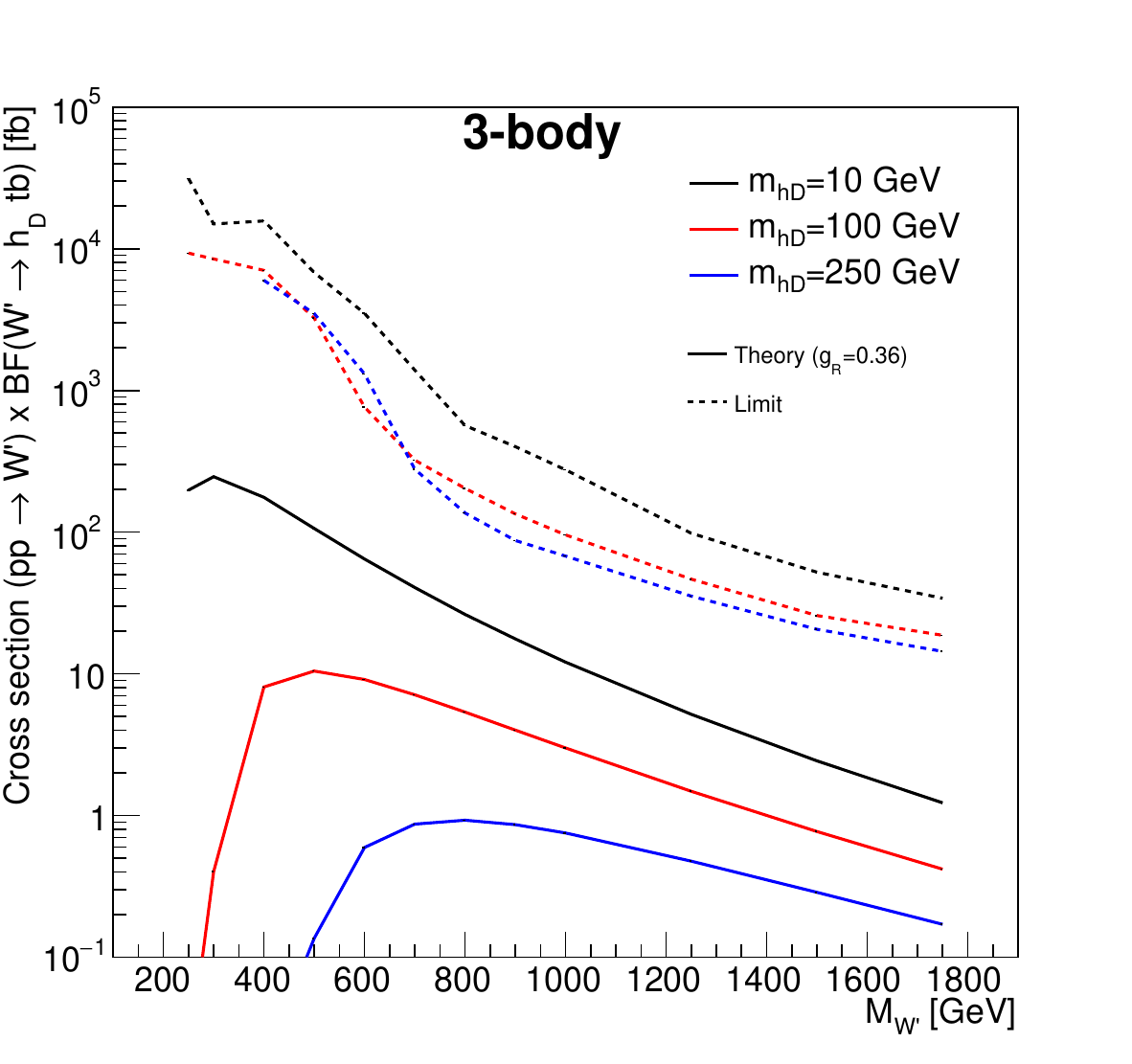}
    \caption{Top: Summary of expected upper limits at 95\% CL on the $h_D W'$ production cross section and the 2-body decay branching fraction of $W'$ as a function of the $W'$ boson mass normalized to an integrated luminosity of 300 fb$^{-1}$ for three choices of the $h_D$ boson mass. Also shown are expected theoretical cross sections and branching fractions at leading order for a coupling value of $g_R=0.36$. Bottom: The same distributions as above except for the 3-body decay of the $W'$ boson.}
    \label{fig:xs_limit_summary}
\end{figure}

\begin{figure}
  \centering
  \includegraphics[width=0.45\textwidth]{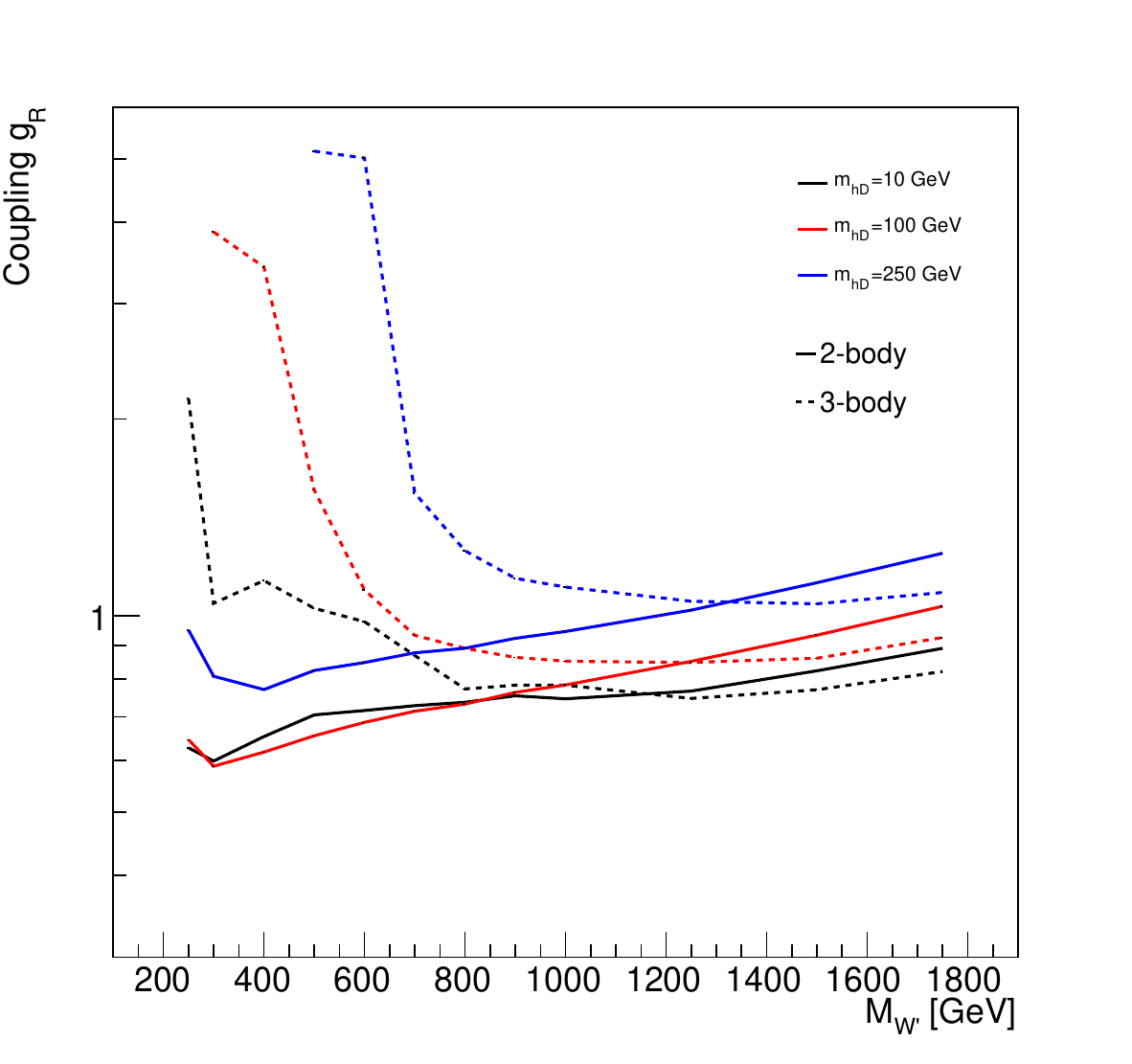}
  \caption{Expected limits on the coupling $(g_{R})$ for an integrated luminosity of 300 fb$^{-1}$ as functions of the $W'$ boson mass for three choices of the $h_D$ boson mass. Results are calculated from the expected limits on the cross section in Fig.~\ref{fig:xs_limit_summary}. }
    \label{fig:g_limit_summary}
\end{figure}

Expected limits are calculated at 95\% CL using a profile likelihood ratio~\cite{Cowan:2010js} with the CLs technique~\cite{Junk:1999kv,Read:2002hq} with {\sc pyhf}~\cite{Feickert:2022lzh,Heinrich:2021gyp} for a binned distribution in the reconstructed mass of the hypothetical $W'$ boson, \rev{with 20 bins}, where bins without simulated background events have been merged into adjacent bins. The background is assumed to have a 50\% relative systematic uncertainty. Expected limits as functions of the $W'$ boson mass are shown in Fig.~\ref{fig:xs_limit_summary} and translated into limits on the coupling ($g_{R}$) in Fig.~\ref{fig:g_limit_summary}.

\section{Discussion}

Studying the 2-body case alone, we find a cross section limit that ranges from $\approx 3~{\rm pb}$, when both the $W'$ boson and the $h_D$ boson are light, down to $\approx 20~{\rm fb}$, when both the $W'$ boson and the $h_D$ boson are relatively heavy. As either the $W'$ boson or the $h_D$ boson becomes more massive, the $\sqrt{\hat{s}}$ of the system is pushed to larger values and leads to better sensitivity.

In the 3-body case, the $\sqrt{\hat{s}}$ of the system only depends on $m_{W'}$. The sensitivity, however, still depends on $m_{h_D}$ through the efficiency, as seen in Fig.~\ref{fig:eff}. The limits here range from $\approx 30~{\rm pb}$ at low mass to $\approx 20~{\rm fb}$ at high mass. Generally, the 2-body and 3-body cases are produced simultaneously and can be considered in a combined limit, however, we make conservative bounds and separate them for the purpose of clarity.

In terms of the left-right model used, Fig.~\ref{fig:g_limit_summary} shows the expected limits on the $SU(2)_R$ gauge coupling $g_R$. At low masses, coupling values are probed down to $\approx 0.6$, while at higher masses, the limits are expected to be marginally weaker. Even though the limits are calculated using a particular left-right model, they will roughly correspond to the limits found in other models with a $W'$ boson. This is because, generically, such models are parameterized by a mass that scales roughly as $\sim g_R v_R$ and a cross section that scales roughly as $\sim g_R^4$.

If DM particles are exclusively produced in the decay of a $h_D$ boson, this model provides a unique opportunity to detect them. In the context of this model, the $h_D$ boson does not couple with any SM particles, making it impossible to produce directly at the LHC, as opposed to other scalar mediators predicted by traditional DM simplified models~\cite{Abdallah:2014hon,Malik:2014ggr}. While the model does include a $Z'$ boson, which could provide an additional signature via \met+$Z'$~\cite{Autran:2015mfa}, here we assume the $Z'$ is too heavy to be visible, making the proposed \met+$W'$ channel the only one accessible at the LHC.

When analyzing the discovery potential for a $W'$ boson in the final state predicted by this model, the presence of significant \met provides a boost to the $W'$ boson, allowing for an increased sensitivity to lower masses when compared to searches for $W'$ bosons produced at rest~\cite{atlascollaboration2023search,cmscollaboration2023search,CMS:2021mux}. A similar argument can be made for the case when the $W'$ boson is boosted by initial state radiation, which is possible in the context of this model. When comparing the \met+$W'$ channel to the jet+$W'$ channel, considerations on the effect of the trigger have to be made. At the LHC, searches that look at final states with hadronic jets, and no other objects, have to rely on datasets that are collected online by triggers, which require high thresholds on the jet transverse momenta. To select a jet+$W'$ final state, a requirement of $\approx 500$ GeV is placed on the recoiling jet $p_T$ to be in trigger efficiency plateau. The \met+$W'$ channel can rely on events selected online by triggers that require large \met. Thresholds are typically lower for those triggers. For example, to select a \met+$W'$ final state, a requirement of $\approx 200$ GeV is place on the \met to be in the trigger efficiency plateau. This provides higher signal efficiency at low mass, and therefore, a stronger limit on the couplings. Previous dedicated searches for $W'\rightarrow tb$ in the low-mass region were performed at the Tevatron, with the results obtained by the CDF experiment in 2015~\cite{CDF:2015aej} still yielding the strongest limits in the 300-900 GeV range\rev{, at the level of $ g_{W'}/g_W < 0.1 (0.2)$ for $m_{W'}=300(500)$ GeV}.

For the simplest case of a right-handed neutrino as dark matter, the decays of $W' \to \ell N_R$ where the $N_R$ is invisible is another relevant search channel~\cite{CMS:2016ifc,ATLAS:2017jbq}.

\section{Conclusions}

In this work we study the search channel of $pp \to W'(tb) h_D$ where the $h_D$ boson decays invisibly. This channel plays a dual role of extending the mono-$X$ program of looking for DM at the LHC through the recoil of the visible object $X$ and extending the searchable range of $W'$ bosons to lower masses. Expanding the mass range for $W'$ boson searches is especially novel and is accomplished through use of \met triggers, which have a lower threshold than comparable hadronic jet triggers.

We estimate that the current LHC dataset could be sensitive to $W'$ boson production in the range from 20 fb to 30 pb, depending on the $W'$ boson mass. These translate to limits on the coupling $(g_R)$ as low as 0.6, which can be interpreted across a fairly generic set of $W'$ boson models. 

Future directions include improved reconstruction algorithms for the $W'$ boson, perhaps using machine learning~\cite{Fenton:2023ikr,Fenton:2020woz} to improve the accuracy of the jet-parton assignment and reconstruction of the missing $z$ component of the invisible decay of the $h_D$ boson.

\section{Acknowledgements}

DW is funded by the DOE Office of Science.
ML is funded by the DOE grant no. DE-SC0007914 and NSF grant no. PHY-2112829. KC is supported in part by the National Science Foundation of China under grant No.~12235001.

\bibliography{wprimedm}
\end{document}